\documentclass[10pt, conference]{IEEEtran}
\usepackage{flushend} 

\usepackage{cite}

\usepackage{float}
\usepackage[colorlinks=true,linkcolor=black,urlcolor=black,citecolor=black]{hyperref}

\usepackage{tikz, tikz-3dplot}
\usetikzlibrary{calc,matrix,positioning,arrows.meta,cd,backgrounds}
\usepackage[american]{circuitikz}

\usepackage{makecell}
\usepackage{multirow}

\usepackage{mathbbol}
\usepackage{amsfonts,amssymb,amsmath,mathtools}
\DeclareSymbolFontAlphabet{\mathbb}{bbold}
\DeclareSymbolFontAlphabet{\mathbbg}{bbold}

\newcommand{\defeq}{:=}

\newcommand{\set}[1]{\ensuremath{\mathbf{#1}}}
\newcommand{\powerset}[1]{\ensuremath{\mathcal{P}(#1)}}

\usepackage{xstring}
\newcommand{\StrHead}[1]{\StrLeft{#1}{1}}
\newcommand{\StrTail}[1]{\StrGobbleLeft{#1}{1}}

\newcommand{\lattice}[1]{\ensuremath{%
	\mathbb{\StrHead{#1}}\mathbf{\StrTail{#1}}
}}
\newcommand{\lat}[1]{\ensuremath{\mathbb{#1}}}
\newcommand{\corr}[1]{R_{\lattice{#1}}}
\newcommand{\abs}[1]{\alpha_{\lattice{#1}}}
\newcommand{\conc}[1]{\gamma_{\lattice{#1}}}
\newcommand{\trans}[2]{\tau_{\mkern 2mu\lattice{#1}}^{\lattice{#2}}}

\newcommand{\lhom}[2]{\phi_{#1}^{#2}}
\newcommand{\opt}[2]{\lattice{#1}_{\set{#2}}}
\newcommand{\ofam}[2]{\lattice{#1}(\set{#2})}
\newcommand{\tol}[2]{\mathbbg{\Lambda}(\lattice{#1},\ofam{#2}{#1})}

\newtheorem{definition}{Definition}[section]
\newtheorem{theorem}{Theorem}[section]

\usepackage{enumerate} 

\usepackage{graphicx}
\makeatletter
\providecommand{\bigsqcap}{%
  \mathop{%
    \mathpalette\updown\bigsqcup%
  }%
}
\newcommand*{\updown}[2]{%
  \rotatebox[origin=c]{180}{$#1#2$}%
}
\makeatother

\title{Formalizing Cyber--Physical System Model Transformation via Abstract Interpretation}
\author{%
	\IEEEauthorblockN{Natasha Jarus, Sahra Sedigh Sarvestani, and Ali Hurson}%
	\IEEEauthorblockA{%
		Department of Electrical and Computer Engineering\\
		Missouri University of Science and Technology\\
		Rolla, USA 65409\\
		Email: \{jarus, sedighs, hurson\}@mst.edu
	}
}

\begin{document}
\maketitle

\begin{abstract}
Model transformation tools assist system designers by reducing the labor--intensive task of creating and updating models of various
aspects of systems, ensuring that modeling assumptions remain consistent across every model of a system, and
identifying constraints on system design imposed by these modeling assumptions.
We have proposed a model transformation approach based on abstract interpretation, a static program analysis technique.
Abstract interpretation allows us to define transformations that are provably correct and specific.
This work develops the foundations of this approach to model transformation.
We define model transformation in terms of abstract interpretation and prove the soundness of our approach.
Furthermore, we develop formalisms useful for encoding model properties.
This work provides a methodology for relating models of different aspects of a system
and for applying modeling techniques from one system domain, such as smart power grids, to other domains, such as water distribution networks.
\end{abstract}

\begin{IEEEkeywords}
Modeling, Model transformation, Formal methods, Abstract interpretation
\end{IEEEkeywords}

\section{Introduction}

The multitude of functional and non--functional requirements for critical infrastructure cyber--physical systems (CPSs) present many challenges to system designers.
A smart grid must be able to supply all its customers; it must be fault--tolerant in the face of component failure; it must be secure against physical and cyber attacks; and it must achieve all these goals with efficient infrastructure.
To meet all these requirements, designers must integrate physical components, cyber control software and hardware, and processes for human operators into a complete system.
This is a truly daunting task, but one that can be facilitated by model-based design and evaluation.

A vast body of literature has been published on various modeling formalisms that capture system performance, dependability, safety, and security.
No single modeling formalism can encompass all aspects of system performance and dependability, necessitating the labor--intensive and error--prone process of creating multiple system models
and propagating  changes across all of these models.
Furthermore, designers must be careful that these models remain consistent with each other, i.e., that the assumptions made about the system by one model are not contradicted by those of any other model.
For instance, a dependability model for a smart grid where two power lines are assumed to be connected in parallel is not compatible with a power flow analysis where the lines are placed in series.

One way to alleviate these challenges is through model transformation, which enables automated or semi-automated transformations between modeling formalisms.
These transformations can ensure that modeling assumptions are consistent across every model of a system by verifying that any model can be transformed into any other.
This approach can also identify constraints on system design imposed by these assumptions.
Such a model transformation approach should meet two design constraints.
First, it should be applicable to a broad range of systems and a variety of modeling formalisms in order to be useful to designers of complex systems.
Second, it should be sound---it should be possible to prove that the result of a transformation is correct and consistent with the initial models.

In our earlier work~\cite{jarus_models_2016}, we proposed a model transformation approach based on \emph{abstract interpretation}, a static program analysis technique.
Models are seen as abstractions of the semantics of a system---its structure and behavior.
Through this lens, provably correct model transformation becomes the problem of defining sound mappings from system semantics to model semantics and vice versa.
By composing these mappings, we can develop sound transformations between modeling formalisms.

The research contribution of this work is twofold. We propose:
\begin{enumerate}
	\item a formalization of system and model semantics, leading to a formalization of sound model transformation, and
	\item a mathematical structure useful in the development of structures that capture system semantics.
\end{enumerate}
Our first contribution formalizes the research approach we outlined in~\cite{jarus_models_2016} and incorporates several improvements from feedback we have received since publication of that work.
The second contribution lays the groundwork for  integrating real--world modeling formalisms into this model transformation approach.

The structure of this paper is as follows: in Section~\ref{sec:related-work}, we briefly summarize related model transformation and formalization techniques.
Section~\ref{sec:abs-int} presents our formalization of system and model semantics and describes how we use this formalization to create a method for model transformation.
Section~\ref{sec:tag-opts} presents tag--option lattices, a structure that we find to be useful when formalizing the semantics of systems.
Section~\ref{sec:conclusion} summarizes our work and discusses future directions for our research.

\section{Related Work}
\label{sec:related-work}

In the literature, \emph{model transformation} refers to two different but related concerns.
One concern is integrating models of different parts of the system into a complete system model; this is more specifically called \emph{ heterogeneous model composition}.
The other concern is transforming one type of model for a system to a different model of the same system or a related system.

Model transformation research specific to CPSs primarily focuses on building hierarchical models~\cite{derler_addressing_2011,feng_scalable_2008,wan_composition_2010,bhave_multidomain_2010}.
Hierarchical models allow different model types to be combined together to model complex systems.
Each component can be modeled in a convenient formalism; the hierarchical model is then simulated by simulating each sub-model in tandem.

The Ptolemy modeling software~\cite{ptolemaeus_system_2014} performs hierarchical modeling and model composition~\cite{xiong_design_2005,lickly_practical_2011}.
As such, Ptolemy makes it easy to build and link small models.
Hierarchical models can consist of heterogeneous sub--models, allowing different parts of the system to be expressed using different types of models~\cite{feng_model_2009,brooks_multimodeling_2008,tripakis_modular_2013,lickly_practical_2011}.
Model composition is achieved in part by defining ontologies of system properties, e.g., units of model inputs and outputs.
Based on these ontologies, Ptolemy can perform conversions of values transmitted between sub--models and check for incompatibilities which indicate modeling errors.
Ptolemy also enables heterogeneous model evaluation: it provides choices for both the modeling language and the solution or simulation technique used to evaluate the model~\cite{goderis_heterogeneous_2009}.
However, Ptolemy does not offer methods for transforming one system--level model to another.
In addition, it is focused on models of system function and lacks facilities for modeling non-functional attributes.

OsMoSys~\cite{vittorini_osmosys_2003} and SIMTHESys~\cite{barbierato_simthesyser_2012} are modeling systems motivated by model--driven engineering.
Their approach to model transformation is based on techniques from software engineering.
Graph--based models, such as Petri Nets and Fault Trees, are described using an object--oriented notation.
Every model has associated interfaces which allow models of different types to be composed and evaluated.
OsMoSys features compositional models and interfaces with external tools to evaluate them~\cite{franceschinis_object_2002}.
SIMTHESys provides a language in which users can describe new modeling formalisms for use with OsMoSys.
Both are capable of modeling both functional and non-functional aspects of a system~\cite{iacono_exploiting_2011}.
However, neither are focused on the problem of model transformation.

M\"obius~\cite{clark_mobius_2001} is another modeling tool that supports hierarchical modeling.
It supports several modeling formalisms, including block diagrams and Petri nets, and additional formalisms including stochastic timed systems can be included via external modeling tools~\cite{gaonkar_performance_2009,buchanan_simulation_2014}.
While this feature offers considerable flexibility in modeling, M\"obius is constructed around a modeling workflow that builds and evaluates hierarchical models and has little support for model transformation.
Its model composition method is based on object--oriented design principles and is applicable to many state--based model formalisms.

AToM$^3$~\cite{delara_atom_2002} is capable of both model transformation and model composition.
It uses metamodels to describe specific modeling languages, then defines transformations between metamodels to transform models~\cite{feng_multiformalism_2007}.
Models are graph-based and transformations take the form of graph rewriting rules~\cite{delara_using_2002}.
However, there is no hierarchy of models, so introducing a new model requires writing transformation rules from the new model to each model that AToM$^3$ implements.

CHESS~\cite{web_chess_2016} provides a modeling language for describing systems and includes several model transformation methods specific to creating dependability models.
CHESS is based on the Unified Modeling Language (UML); transformations are based on graph rewriting rules.
CONCERTO~\cite{web_concerto_2016} extends CHESS by introducing modeling techniques for non-functional system attributes such as dependability~\cite{montecchi_reusable_2013}.
However, CONCERTO is focused on multicore processing systems~\cite{bonfiglio_executable_2015,de_matos_pedro_towards_2014} and lacks the features necessary for modeling complex physical components.

Rosetta~\cite{kong_rosetta_2003} is focused on functional multi--formalism modeling~\cite{streb_using_2006}.
It takes an algebraic approach to relating models: each formalism is described as a coalgebra---a mathematical system useful for describing arbitrary transitions among arbitrary states~\cite{frisby_model_2011,alexander_rosetta_2009}.
The coalgebras corresponding to each formalism are placed in a lattice, which provides a structure for determining how to transform one model into another.
Model transformations can be used to relate different models of the same system; for example, it is possible to combine a functional system model with a model of that system's power consumption.
However, Rosetta lacks many features required for CPS modeling, especially support for hybrid discrete--continuous formalisms.

Each of these model transformation tools offers a partial solution to the model transformation problem; however, none of them present a solution that is generally applicable.
Some frameworks place constraints on the behavior of transformation functions (e.g., class inheritance transformation).
Others apply only to specific formalisms.
Furthermore, only Rosetta offers an approach that can  be proven to be correct.
The work presented in this paper aims to address these shortcomings by providing a model transformation approach that relates a wide variety of modeling formalisms in a provably sound fashion, and yields results that are sufficiently specific to be meaningful.

\section{Abstract Interpretation of Models}
\label{sec:abs-int}

The foundation of our approach is abstract interpretation~\cite{cousot_abstract_1977,nielson_principles_1999}, a formalism for developing sound semantic abstractions.
In this work, system semantics are represented in terms of properties that hold for the system.
Such properties might include information about components, their reliabilities, and how they are interconnected.
Models are abstractions of system semantics---they concern certain properties of the system, but not others.
Thus, generating a model from a system's properties, then deriving properties of that system from the generated model, may result in some of the initial properties not being present in the derived properties.
This is a necessary effect of abstraction---we cannot derive properties from a model if the model does not capture those properties.
To define mappings from system properties to models and vice versa, both domains need to allow for this potential loss of precision.

\subsection{Properties}

We first define how system semantics are represented.
\emph{Lattices} (see~\cite{davey_priestley_2002}) 
offer a useful formalism for describing the nature of approximation.
We  define a complete $\lattice{Properties}$ lattice ordered by specificity: for $p_1, p_2 \in \lattice{Properties}, p_1 \sqsubseteq p_2$ means that the constraints in $p_1$ and $p_2$ are not contradictory and that $p_1$ places the same or more constraints on a system than $p_2$ does.
For example, $p_2$ could constrain the reliability of a component to fall in the range $(0,1]$, whereas $p_1$ could require that component to have a reliability of $0.95$.

The \emph{meet} (denoted as $\sqcap$) of two elements of $\lattice{Properties}$ places the constraints of both elements on a system; the \emph{join} (denoted as $\sqcup$) implies satisfaction of the constraints of either element.
Suppose $p_1$ requires a component's reliability to fall in $[0.8,1.0]$ and $p_2$ constrains it within $[0.75,0.9]$.
Then $p_1 \sqcap p_2$ will require it to be in $[0.8,0.9]$ and $p_1 \sqcup p_2$ within $[0.75,1.0]$.
$\bigsqcap$ and $\bigsqcup$ extend this concept to subsets of $\lattice{Properties}$. 

For certain $p_1, p_2 \in \lattice{Properties}$ are contradictory,  $p_1 \sqcap p_2$ will result in a constraint that is impossible to satisfy.
If $p_1$ requires a component to have a reliability in $[0.5,0.7]$ and $p_2$ requires it in $[0.9,1]$, then it is impossible for any component to meet both constraints.
In this paper, we require that every element of $\lattice{Properties}$ to be satisfiable except for $\bot$, the ``impossible'' constraint.
Therefore, for this example, $p_1 \sqcap p_2 = \bot$. Note that $\forall p \in \lattice{Properties}, \bot \sqsubseteq p$.

To summarize, each element of the $\lattice{Properties}$  lattice describes one or more systems. In the general case, $p$ describes a set of systems, all of which meet the constraints in $p$.
If every constraint in $p \in \lattice{Properties}$ has exactly one possible choice, $p$ will describe a single system.

\subsection{Models}

We now consider how modeling formalisms can be represented in this lattice framework.
As a given element of the $\lattice{Properties}$ lattice may not define a single system, we must account for the possibility that the lattice may not specify the system well enough for a single model to be abstracted from it.
If $p \in \lattice{Properties}$ does not constrain the reliability of a component to a single value, a single reliability model cannot be abstracted from $p$.
Instead, we abstract a set of models, one for each possible assignment of the component's reliability, subject to the constraints of $p$.

Therefore, in the same way that the $\lattice{Properties}$ lattice is defined, we also define the domain of each modeling formalism to account for the nature of potentially imprecise system specifications.
To ensure that this approach is broadly applicable, we define this domain using structure external to the modeling formalism itself.
Thus we do not have to require, say, that a reliability model formalism be able to express the concept of a component having a range of possible reliabilities.

We use a \emph{powerset lattice} to provide this extra structure.
For a given model formalism, the set $\set{Model}$ contains all possible models expressible in that formalism.
The powerset lattice $\powerset{\set{Model}}$ then forms a lattice ordered by specificity: for $M_1, M_2 \subseteq \set{Model}$, $M_1 \subseteq M_2$ indicates that $M_1$ contains fewer possible models describing a system, and thus places more constraints on the system, than $M_2$ does.
Likewise, $M_1 \cap M_2$ produces a set of models that fit the constraints associated with $M_1$ and with $M_2$; $M_1 \cup M_2$ produces a set of models where constraints from either may hold.

Singleton sets (i.e., sets of the form $\{m\}$, $m \in \set{Model}$) correspond to fully-specified models, and $\emptyset = \bot$ corresponds to an ``impossible'' system---one with contradictory modeling requirements.

To make the notation clearer and more consistent, we will define $\lattice{Model} = \powerset{\set{Model}}$, as the powerset lattice of the original set of models,  $\set{Model}$ .
For the powerset lattice $\lattice{Model}$, we will use the rounded operators ($\subseteq, \bigcap, \bigcup$) to prevent confusion with the square operators of the lattice $\lattice{Properties}$, and of lattices in the abstract.

\subsection{Correctness}

In this work, we represent the set of systems by $\mathcal{S}$.
We think of these systems abstractly; thus, we do not concern ourselves with the representation of $\mathcal{S}$ or its elements.
When we speak of a system $s \in \mathcal{S}$, we understand $s$ to be the system to be modeled.

Any system $s \in \mathcal{S}$ is described by a number of elements of $\lattice{Properties}$.
To formalize this notion, we use a \emph{correctness relation} to relate a system to properties (and later, models) that describe it.
We suppose a relation $\corr{P} : \mathcal{S} \rightarrow \lattice{Properties}$ where $s\,\corr{P}\,p$ if and only if $p$ describes the system $s$.
We must assume the existence of $\corr{P}$, since the properties of the system being designed are determined by the designer.
However, abstract interpretation allows us to induce correctness relationships between systems and models based on $\corr{P}$---in other words,
abstract interpretation enables sound transformations between system properties and system models.

\begin{definition}
	\label{defn:correctness-rel}
	A \emph{correctness relation} $\corr{L} : \mathcal{S} \rightarrow \lattice{L}$ relates systems to elements of a lattice $\lattice{L}$.
	Two attributes hold for $\corr{L}$:
	\begin{enumerate}[(i)]
		\item \label{prop:more-correct}
			If $s\,\corr{L}\,l_1$ and $l_1 \sqsubseteq l_2$, then $s\,\corr{L}\,l_2$.
		\item \label{prop:most-correct}
			If $\forall l \in \set{L'} \subseteq \lattice{L}, s\,\corr{L}\,l$, then $s\,\corr{L}\,\bigsqcap\set{L'}$.
	\end{enumerate}
\end{definition}

In terms of $\lattice{Properties}$ and its correctness relation $\corr{P}$,
Property~(\ref{prop:more-correct}) states that we can relax correct constraints without violating their correctness. The reverse does not hold, otherwise, the inconsistent constraint $\bot$ would describe every system.
The formalization of relaxation of constraints as described by Property~(\ref{prop:more-correct})  allows us to generalize constraints and therefore plays a crucial role in modeling abstraction.

Property~(\ref{prop:most-correct}) requires that for any set of constraints $\set{L'}$ there exist a ``best'' constraint that correctly describes any system described by every constraint in $\set{L'}$.
We can apply this property to the constraints derived from several models to narrow down our description of a given system's properties.
In this sense, it allows us to derive a specific result from a number of more general results.
Note that the converse of (\ref{prop:most-correct}) follows from (\ref{prop:more-correct}), so (\ref{prop:most-correct}) could also be written as a biconditional.

\vspace{-.1em}
\subsection{Abstraction and Concretization}

Given a correctness relation $\corr{P}$ for $\lattice{Properties}$, we desire to define a mapping between $\lattice{Properties}$ and a modeling formalism $\lattice{Model}$ that induces a correctness relation $\corr{M} : \mathcal{S} \rightarrow \lattice{Model}$.
Furthermore, this mapping must allow for the modeling domain to abstract system constraints.
For instance, a topology model should be able to discard constraints on component reliability.

The formalism of choice for this task is a \emph{Galois connection}:
\begin{definition}
	A \emph{Galois Connection} $(\lattice{P}, \alpha, \gamma, \lattice{M})$ between two complete lattices $\lattice{P}$ and $\lattice{M}$ consists of a pair of monotone functions $\alpha : \lattice{P} \rightarrow \lattice{M}$ and $\gamma :  \lattice{M} \rightarrow \lattice{P}$ for which the following relationships hold:
	\begin{align}
		(\gamma \circ \alpha)(p) &\sqsupseteq p
		\label{eqn:abs-conc} \\
		(\alpha \circ \gamma)(m) &\sqsubseteq m
		\label{eqn:conc-abs}
	\end{align}

	We refer to $\lattice{P}$ as the \emph{concrete domain}, $\lattice{M}$ as the \emph{abstract domain}, $\alpha$ as the \emph{abstraction operator}, and $\gamma$ as the \emph{concretization operator}.
\end{definition}

In terms of models and properties, $\alpha$ abstracts a model,  $m$,  from a set of constraints on a system,  $p$,  and $\gamma$ derives, or \emph{concretizes}, system constraints from a model of that system.
Relationship (\ref{eqn:abs-conc}) states that abstracting the model  $m$ from constraints $p$, then concretizing constraints from that model, results in constraints that are at most more general than those of $p$.
In other words, abstraction may relax constraints irrelevant to the model formalism, but it cannot produce a model that implies constraints that contradict $p$.
Relationship (\ref{eqn:conc-abs}) requires that $\lattice{Properties}$ be able to completely capture the constraints imposed by each model formalism, meaning that if constraints are concretized from a model,  $m$, of a system, any other model abstracted from these constraints will be as least as specific as the original model,  $m$. Concretization may introduce additional constraints, but in practice, the $\sqsubseteq$ of (\ref{eqn:conc-abs}) will often be strict equality in practice.

Next, we show that each Galois connection induces a correctness relation $\corr{M}$ on the abstract domain.

\begin{theorem}
	Given a Galois connection $(\lattice{P}, \alpha, \gamma, \lattice{M})$ and a correctness relation $\corr{P} : \mathcal{S} \rightarrow \lattice{P}$, the relation $\corr{M} : \mathcal{S} \rightarrow \lattice{M}$ defined by $s\,\corr{M}\,m \iff s\,\corr{P}\,\gamma(m)$ is a correctness relation.
\end{theorem}
\begin{IEEEproof}
	We must show that properties (\ref{prop:more-correct}) and (\ref{prop:most-correct}) from Definition~\ref{defn:correctness-rel} hold for $\corr{M}$.
	Take $s \in \mathcal{S}$ and $m_1, m_2 \in \lattice{M}$.
	\begin{align*}
		  & s\,\corr{M}\,m_1 \wedge m_1 \sqsubseteq m_2
		  \\
		\iff
		  & s\,\corr{P}\,\gamma(m_1) \wedge m_1 \sqsubseteq m_2
		  & (\text{Defn. of $\corr{M}$})
		  \\
		\iff
		  & s\,\corr{P}\,\gamma(m_1) \wedge \gamma(m_1) \sqsubseteq \gamma(m_2)
		  & (\text{$\gamma$ monotone})
		  \\
		\implies
		  & s\,\corr{P}\,\gamma(m_2)
		  & (\text{Prop. (\ref{prop:more-correct}) for $\corr{P}$})
		  \\
		\iff
		  & s\,\corr{M}\,m_2
		  & (\text{Defn. of $\corr{M}$})
	\end{align*}

	The proof of (\ref{prop:most-correct}) uses the fact that $\gamma$ is completely multiplicative, that is, $\bigsqcap\{\gamma(m) \mid m \in \set{M'}\} = \gamma\left(\bigsqcap\set{M'}\right)$.
	Take $s \in \mathcal{S}$ and $\set{M'} \subseteq \lattice{M}$.
	\begin{align*}
		  & \forall m \in \set{M'} s\,\corr{M}\,m \\
		\iff
		  & \forall m \in \set{M'}, s\,\corr{P}\,\gamma(m)
		  & (\text{Defn. of $\corr{M}$})
		  \\
		\implies
		  & s\,\corr{P}\,\bigsqcap\left\{\gamma(m) \mid m \in \set{M'}\right\}
		  & (\text{Prop. (\ref{prop:most-correct}) for $\corr{P}$})
		  \\
		\iff
		  & s\,\corr{P}\,\gamma\left(\bigsqcap\set{M'}\right)
		  & (\text{Multiplicativity of $\gamma$})
		  \\
		\iff
		  & s\,\corr{M}\,\bigsqcap\set{M'}
		  & (\text{Defn. of $\corr{M}$})
	\end{align*}
\end{IEEEproof}

Put in terms of models and system properties, if we define a Galois connection between $\lattice{Properties}$ and the lattice for a given modeling formalism $\lattice{Model}$, then every correct collection of system constraints abstracts to a correct model and every correct model concretizes to a correct collection of system constraints.
Therefore, we have developed a provably sound definition of the nature of model abstraction.


\subsection{Model Transformation}

Given this formalization of system and model semantics, we can now formalize the problem of model transformation.
Suppose we have a properties domain and two modeling formalisms with associated Galois connections to the properties domain
$(\lattice{Properties},\abs{M_1}, \conc{M_1}, \lattice{Model_1}$) and
$(\lattice{Properties},\abs{M_2}, \conc{M_2}, \lattice{Model_2}$).
Furthermore, we have a correctness relation $\corr{P}$ which induces correctness relations $\corr{M_1}$ and $\corr{M_2}$.

\begin{definition}
	A \emph{model transformation} from $\lattice{Model_1}$ to $\lattice{Model_2}$ is a semantically sound mapping $\trans{M_1}{M_2} : \lattice{Model_1} \rightarrow \lattice{Model_2}$.
	That is, if $m_1 \in \lattice{Model_1}$ is correct, then $\trans{M_1}{M_2}(m_1)$ is also correct.
\end{definition}

We can define $\trans{M_1}{M_2}$ by first concretizing constraints from $m_1 \in \lattice{Model_1}$, then abstracting an element of $\lattice{Model_2}$ from it.

\begin{theorem}
	The mapping $\trans{M_1}{M_2}(m_1) = (\abs{M_2} \circ \conc{M_1})(m_1)$ is sound.
\end{theorem}
\begin{IEEEproof}
	Take $s \in \mathcal{S}$ and $m_1 \in \lattice{Model_1}$.
	\begin{align*}
		  & s\,\corr{M_1}\,m_1
		  \\
		\iff
		  & s\,\corr{P}\,\conc{M_1}(m_1)
		  & \text{Defn. of $\corr{M_1}$}
		  \\
		\implies
		  & s\,\corr{P}\,(\conc{M_2} \circ \abs{M_2} \circ \conc{M_1})(m_1)
		  & \text{Eqn. (\ref{eqn:abs-conc}), Prop. (\ref{prop:more-correct})}
		  \\
		\iff
		  & s\,\corr{M_2}\,(\abs{M_2} \circ \conc{M_1})(m_1)
		  & \text{Defn. of $\corr{M_2}$}
	\end{align*}
\end{IEEEproof}

To sum up the transformation process: begin with a model $m_1 \in \set{Model_1}$.
Concretize properties of the system from $\{m_1\}$, then apply $\trans{M_1}{M_2}$ to produce a set of models $M'_2 \subseteq \set{Model_2}$.
Finally, select a model from $M'_2$ by introducing information about the system not present in $m_1$.

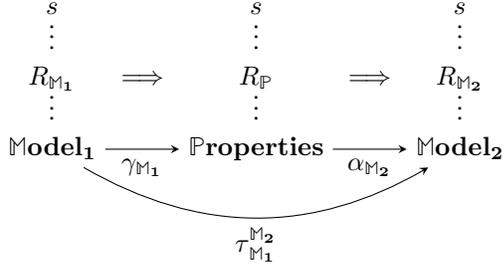
\begin{figure}[!ht]
\centering
\begin{tikzpicture}
	\matrix(m) [matrix of math nodes, row sep=0em, column sep=0em, minimum width=1em, minimum height=1em, text depth=.25ex, text height=1.25ex]
	{%
		s & & s & & s\\
		\vdots & & \vdots & & \vdots \\
		\corr{M_1} & \implies & \corr{P} & \implies & \corr{M_2}\\
		\vdots & & \vdots & & \vdots \\
		\lattice{Model_1} & & \lattice{Properties} & & \lattice{Model_2}\\
	};

	\path[-stealth]
	(m-5-1) edge node [below] {$\conc{M_1}$} (m-5-3)
	(m-5-3) edge node [below] {$\abs{M_2}$}  (m-5-5)
	(m-5-1) edge [bend right] node [below] {$\trans{M_1}{M_2}$} (m-5-5)
	;
\end{tikzpicture}
\caption{Sound model transformation.}
\label{fig:model-trans}
\end{figure}

Figure~\ref{fig:model-trans} illustrates the domains, mappings, and relationships present in this formalization of model transformation.

\subsection{Selection and Specificity}

Recall that the elements of $\lattice{Model}$ are sets of models.
To concretize properties of a single model $m \in \set{Model}$, we first map it to $\{m\} \in \lattice{Model}$, then apply $\gamma$.
Conversely, for a set of models $M$ produced from an abstraction operation, each model in that set equally captures the system constraints from which $M$ was abstracted.
If $M = \emptyset$, then the chosen modeling formalism cannot reason about the given system constraints.
If $M = \{m\}$, then the abstraction process has produced a single model describing the system.
Otherwise, the system constraints lack some information about the system that is relevant to this modeling formalism.
In this case, the user must introduce new information about the system by selecting one model from this set.
For example, one may have to provide information about component reliability when selecting a reliability model.

We represent this selection process as a function $\sigma : \lattice{Model} \rightarrow \set{Model}$; the definition of $\sigma$ depends entirely upon the exact system being modeled.
While the known system constraints may not be precise enough to indicate exactly which model in the set is correct,
they still indicate that the correct model is in the given set of models.
Therefore we can constrain $\sigma$ to not produce a model which we know is incorrect even when we do not have enough information about the system to produce a single model.

\begin{definition}
	The function $\sigma : \lattice{Model} \rightarrow \set{Model}$ is a \emph{selection operator} if the following conditions hold:
	\begin{enumerate}[(i)]
		\item $\sigma(m) \in m$
		\item If $s\,\corr{M}\,m$, then $s\,\corr{M}\,\{\sigma(m)\}$
	\end{enumerate}
\end{definition}

Given a selection operator, we can incorporate the newly introduced information back into the system properties domain, allowing future transformations to include these constraints and therefore produce more specific results.
Take $p \in \lattice{Properties}$ such that $s\,\corr{P}\,p$.
Derive the exact model of formalism $\lattice{Model}$ by letting $m \defeq (\sigma \circ \alpha)(p)$.
By definition of $\sigma$ we know $s\,\corr{M}\,\{m\}$, so $s\,\corr{P}\,\gamma(\{m\})$.
Finally, we can construct a more specific element $p' \in \lattice{Properties}$ by $p' \defeq p \sqcap \gamma(\{m\})$.
The correctness of $p'$ follows from property~(\ref{prop:most-correct}) for $\corr{P}$, and by definition of $\sqcap$, $p' \sqsubseteq p$.

\begin{figure}[!ht]
\centering
\begin{tikzpicture}
	\matrix(m) [matrix of math nodes, row sep=6em, column sep=8em, minimum width=2em, minimum height=2em, text depth=.25ex, text height=1.25ex]
	{%
		\lattice{Model} & \lattice{Properties} \\
		\set{Model}\\
	};

	\path[-stealth]
	(m-1-2) edge [transform canvas={yshift= 1mm}] node [above] {$\alpha$} (m-1-1)
	(m-1-1) edge [transform canvas={yshift=-1mm}] node [below] {$\gamma$} (m-1-2)
	(m-1-1) edge [transform canvas={xshift=-1mm}] node [left] {$\sigma$} (m-2-1)
	(m-2-1) edge [transform canvas={xshift= 1mm}] node [right] {$m \mapsto \{m\}$} (m-1-1)
	;
\end{tikzpicture}
	\caption{Mappings between $\set{Model}$,
	$\mathbb{M}\mathbf{odel}$, and $\mathbb{P}\mathbf{roperties}$}
\label{fig:model_property_diagram}
\end{figure}
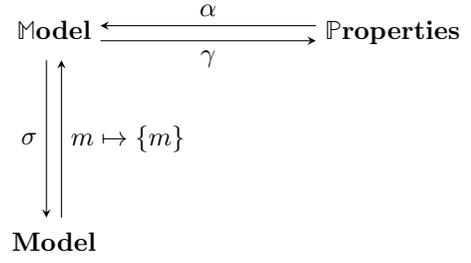

Figure~\ref{fig:model_property_diagram} depicts the relationship between these given functions and domains.

\section{Tag-Options Lattice}
\label{sec:tag-opts}

In our formalization of systems and models, we assume a properties domain that is a lattice of constraints on a system; its elements are ordered by specificity.
A common pattern arises when defining this domain: a lattice that assigns a set of potential values to each element of a set of names or tags.
Two examples are assigning possible reliabilities to components and defining whether a state is considered functional or failed.
We will refer to this type of lattice as a \emph{Tag--Options Lattice}; it is comprised of a \emph{tag lattice} and a family of \emph{options lattices}.
For instance, one may use a tag lattice where each element is a set of components known to be part of a system; each element of the corresponding options lattice is a function that assigns possible reliabilities to each component.

\subsection{Tag Lattice}
Let $\set{T} \defeq \{t_1,t_2,\cdots\}$ be a set of tags.
\begin{definition}
	The \emph{tag lattice} $\lattice{T} \defeq \powerset{\set{T}}^{\partial}$ is the dual of the powerset lattice of $\set{T}$, where $\sqsubseteq \defeq \supseteq$, $\bigsqcap \defeq \bigcup$, and $\bigsqcup \defeq \bigcap$.
\end{definition}

In this lattice, $\top = \emptyset$ corresponds to a system where no tags are known to apply---for example, a system with no known components.
Thus, every system is described by $\top$.
Lattice elements are ordered by specificity; if $T_1,T_2 \in \lattice{T}$ and $T_1 \sqsubseteq T_2$, then $T_1$ contains more information than $T_2$ about tags that apply to a system.


\subsection{Options Lattice Family}

Let $\set{O} = \{o_1,o_2,\cdots\}$ denote the set of options---potential values---for each tag.

\begin{definition}
	For each set of tags $\set{T'} \in \lattice{T}$ we can define a corresponding \emph{options lattice} $\opt{O}{T'}$.
	The elements of $\opt{O}{T'}$ are functions $f : \set{T'} \rightarrow \powerset{\set{O}}$ that assign a set of possible options to each tag.
	For any $f,g \in \opt{O}{T'}$, $f \sqsubseteq g$ if and only if $f(t) \subseteq g(t)$ for all $t \in \set{T'}$.

	We can alternatively view the elements as sets of tuples $(t,\set{o})$ where $t \in \set{T'}$ and $\set{o} \in \powerset{\set{O}}$.
	Each set contains exactly one tuple per tag.

	For any set of elements of an options lattice, $\set{O'} \subseteq \opt{O}{T'}$, we define
	\begin{enumerate}[i)]
		\item $\bigsqcup \set{O'} = \lambda t.\bigcup \{ f(t) \mid f \in \set{O'}\}$, and
		\item $\bigsqcap \set{O'} = \lambda t.\bigcap \{ f(t) \mid f \in \set{O'}\}$.
	\end{enumerate}

	We refer to the \emph{family of options lattices} associated with tag set $\set{T}$ by $\ofam{O}{T}$.
\end{definition}

For example, if $\set{T'} \defeq \{t_1, t_2\}$ and $\set{O} \defeq \{x,y\}$, then
$f \defeq \{(t_1, \emptyset), (t_2, \{x,y\})\}$ and $g \defeq \{(t_1, \{x\}), (t_2, \{y\})\}$ are elements of $\opt{O}{T'}$.
Furthermore,
\begin{align*}
	f \sqcup g &= \{(t_1, \{x\}), (t_2, \{x,y\})\} \\
	f \sqcap g &= \{(t_1, \emptyset), (t_2, \{y\})\}
\end{align*}

\begin{theorem}
	$\opt{O}{T'}$ is a complete lattice.
\end{theorem}
\begin{IEEEproof}
	The proof that $\sqsubseteq$ is a partial order on $\opt{O}{T'}$ follows directly from $\subseteq$ being a partial order on $f(t), f \in \opt{O}{T'}$ for all $t \in \set{T'}$. Likewise, the proof that $\bigsqcap$ and $\bigsqcup$ are complete follows from the completeness $\bigcap$ and $\bigcup$.
\end{IEEEproof}

From these definitions it follows that $\top(t) = \set{O}$ and $\bot(t) = \emptyset$ for all $t \in \set{T'}$.
We can always imagine a system where any of the given options holds for each tag; a system where no tag corresponds to any of the options is a system about which our abstractions cannot reason.

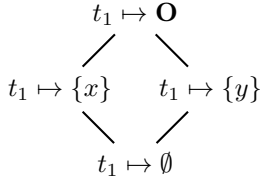
\begin{figure}
	\centering
	\begin{tikzpicture}
		\node (nO) at (0,0) {$t_1 \mapsto \set{O}$};
		\node (nx) at (-1,-1) {$t_1 \mapsto \{x\}$};
		\node (ny) at (1,-1) {$t_1 \mapsto \{y\}$};
		\node (n0) at (0,-2) {$t_1 \mapsto \emptyset$};

		\draw [thick] (nO) -- (nx) -- (n0) -- (ny) -- (nO);
	\end{tikzpicture}
	\caption{Hasse diagram for $\lat{O}_{\{t_1\}}$}
	\label{fig:1-elt-hasse}

\end{figure}

\begin{figure}
	\centering
	\tdplotsetmaincoords{55}{135}

	\begin{tikzpicture}[scale=0.9,tdplot_main_coords, every node/.style={fill opacity=0.5,fill=white,text opacity=1,align=left},font=\small]
		\node (ROO) at (0, 0, 3) {$t_1 \mapsto \set{O}$ \\ $t_2 \mapsto \set{O}$};

		\node (RxO) at (3,0,-1) {$t_1 \mapsto \{x\}$ \\ $t_2 \mapsto \set{O}$};
		\node (ROx) at (0,-3,-1) {$t_1 \mapsto \set{O}$ \\ $t_2 \mapsto \{x\}$};
		\node (RyO) at (-3,0,-1) {$t_1 \mapsto \{y\}$ \\ $t_2 \mapsto \set{O}$};
		\node (ROy) at (0,3,-1) {$t_1 \mapsto \set{O}$ \\ $t_2 \mapsto \{y\}$};

		\node (Rxx) at (3,-3,-4) {$t_1 \mapsto \{x\}$ \\ $t_2 \mapsto \{x\}$};
		\node (Rxy) at (3,3,-4) {$t_1 \mapsto \{x\}$ \\ $t_2 \mapsto \{y\}$};

		\node (Ryx) at (-3,-3,-4) {$t_1 \mapsto \{y\}$ \\ $t_2 \mapsto \{x\}$};
		\node (Ryy) at (-3,3,-4) {$t_1 \mapsto \{y\}$ \\ $t_2 \mapsto \{y\}$};

		\node (Rx0) at (3,0,-7) {$t_1 \mapsto \{x\}$ \\ $t_2 \mapsto \emptyset$};
		\node (R0x) at (0,-3,-7) {$t_1 \mapsto \emptyset$ \\ $t_2 \mapsto \{x\}$};

		\node (Ry0) at (-3,0,-7) {$t_1 \mapsto \{y\}$ \\ $t_2 \mapsto \emptyset$};
		\node (R0y) at (0,3,-7) {$t_1 \mapsto \emptyset$ \\ $t_2 \mapsto \{y\}$};

		\node (R0O) at (0.5,-0.5,-4.5) {$t_1 \mapsto \emptyset$ \\ $t_2 \mapsto \set{O}$};
		\node (RO0) at (-0.5,0.5,-3.5) {$t_1 \mapsto \set{O}$ \\ $t_2 \mapsto \emptyset$};

		\node (R00) at (0,0,-11) {$t_1 \mapsto \emptyset$ \\ $t_2 \mapsto \emptyset$};

		\begin{scope}[on background layer]
		\draw [orange,thick] (ROx) -- (Rxx) -- (R0x) -- (Ryx) -- (ROx);
		\draw [purple,thick] (RxO) -- (Rxx) -- (Rx0) -- (Rxy) -- (RxO);

		\draw [blue,thick] (RyO) -- (Ryx) -- (Ry0) -- (Ryy) -- (RyO);
		\draw [green!60!black,thick] (ROy) -- (Rxy) -- (R0y) -- (Ryy) -- (ROy);

		\draw [brown,thick] (ROO) -- (RxO) -- (R0O) -- (RyO) -- (ROO);
		\draw [violet,thick] (ROO) -- (ROx) -- (RO0) -- (ROy) -- (ROO);

		\draw [black,thick] (R0O) -- (R0x) -- (R00) -- (R0y) -- (R0O);
		\draw [red,thick] (RO0) -- (Rx0) -- (R00) -- (Ry0) -- (RO0);
		\end{scope}

	\end{tikzpicture}

	\caption{Hasse diagram for $\lat{O}_{\{t_1,t_2\}}$}
	\label{fig:2-elt-hasse}
\end{figure}
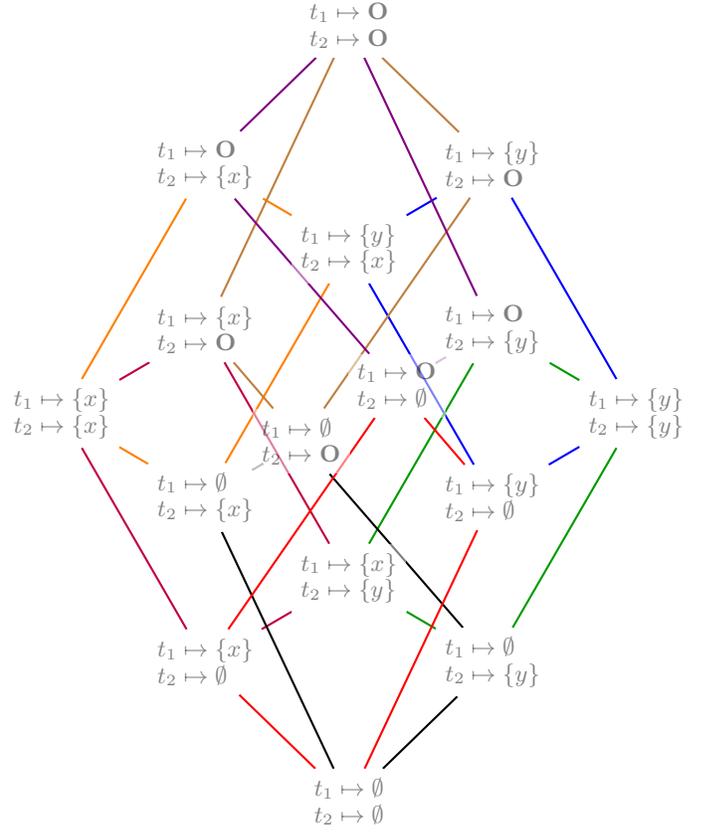

Hasse diagrams for $\lattice{O}_{\{t_1\}}$ and $\lattice{O}_{\{t_1,t_2\}}$ are shown in Figure~\ref{fig:1-elt-hasse} and Figure~\ref{fig:2-elt-hasse}, respectively. 

%
%

Thus far, we have defined a lattice of system tags $\lattice{T}$ and a family of options lattices $\lattice{O}(\set{T}) \defeq \{\opt{O}{T'} \mid \set{T'} \in \lattice{T}\}$ consisting of mappings of tags to options.
What remains is to combine these lattices into a single tag--options lattice.

\subsection{Options Lattice Homomorphisms}

Before we can develop a tag--options lattice, we must define how elements of different options lattices are related.
The tool of choice is a \emph{lattice homomorphism}: a mapping between lattices that preserves meets and joins.

\begin{definition}
	For all sets of tags $\set{A},\set{B} \in \lattice{T}$
	we define a function $\lhom{A}{B} : \opt{O}{A} \rightarrow \opt{O}{B}$ by
	\begin{equation*}
		\lhom{A}{B}(f) \defeq \lambda t.
		\begin{cases}
			f(t) & \text{if } t \in \set{A} \cap \set{B} \\
			\set{O} & \text{if } t \in \set{B} - \set{A}
		\end{cases}
	\end{equation*}

	Or, from a sets-of-tuples perspective,
	\begin{equation*}
		\lhom{A}{B}(f) = \{(t,\set{o}) \in f \mid t \in \set{A} \cap \set{B}\} \cup \{(t,\set{O}) \mid t \in \set{B} - \set{A}\}
	\end{equation*}
\end{definition}

$\phi$ allows us to convert a function with one domain to a related function with a different domain: if $f: \set{A} \rightarrow \powerset{\set{O}}$, then $\lhom{A}{B}(f) : \set{B} \rightarrow \powerset{\set{O}}$.

\begin{theorem}
	$\lhom{A}{B}$ is a lattice homomorphism. That is, for all $\set{O'} \subseteq \opt{O}{A}$,
	\begin{enumerate}[i)]
		\item $\bigsqcap \left\{\lhom{A}{B}(f) \mid f \in \set{O'}\right\} = \lhom{A}{B}\left(\bigsqcap \set{O'}\right)$ and
		\item $\bigsqcup \left\{\lhom{A}{B}(f) \mid f \in \set{O'}\right\} = \lhom{A}{B}\left(\bigsqcup \set{O'}\right)$.
	\end{enumerate}
\end{theorem}

\begin{IEEEproof}
	To show $\bigsqcap \left\{\lhom{A}{B}(f) \mid f \in \set{O'}\right\} = \lhom{A}{B}\left(\bigsqcap \set{O'}\right)$,
	suppose $\set{A},\set{B} \subseteq \set{T}$ and $\set{O'} \subseteq \opt{O}{A}$ and
	take arbitrary $t \in \set{B}$.

	\noindent \emph{Case 1}: $t \in \set{A}$.
	\begin{align*}
		&\lhom{A}{B}\left(\bigsqcap \set{O'}\right)(t) \\
		=& \left(\bigsqcap \set{O'}\right)(t) & (\text{Defn.\ of~}\lhom{A}{B})\\
		=& \bigcap \left\{f(t) \mid f \in \set{O'}\right\} & (\text{Defn.\ of~} \bigsqcap \text{~for~} \set{A})\\
		=& \bigcap \left\{\lhom{A}{B}(f)(t) \mid f \in \set{O'}\right\} & (\text{Defn.\ of~}\lhom{A}{B})\\
		=& \bigsqcap \left\{\lhom{A}{B}(f) \mid f \in \set{O'}\right\}(t) & (\text{Defn.\ of~} \bigsqcap \text{~for~} \set{B})
	\end{align*}

	\noindent\emph{Case 2:} $t \notin \set{A}$.
	\begin{align*}
		& \lhom{A}{B}\left(\bigsqcap \set{O'}\right)(t) \\
		=& \set{O} & (\text{Defn.\ of~}\lhom{A}{B})\\
		=& \bigcap \left\{\set{O} \mid f \in \set{O'}\right\} & (\text{Set Properties})\\
		=& \bigcap \left\{\lhom{A}{B}(f)(t) \mid f \in \set{O'}\right\} & (\text{Defn.\ of~}\lhom{A}{B})\\
		=& \bigsqcap \left\{\lhom{A}{B}(f) \mid f \in \set{O'}\right\}(t) & (\text{Defn.\ of~} \bigsqcap \text{~for~} \set{B})
	\end{align*}

	The proof that $\bigsqcup \left\{\lhom{A}{B}(f) \mid f \in \set{O'}\right\} = \lhom{A}{B}\left(\bigsqcup \set{O'}\right)$ is analogous.
\end{IEEEproof}

We now prove a corollary necessary to show the completeness of the tag--option lattice:

\begin{theorem}
	\label{thm:lhom-order-corr}
	If $\set{A},\set{B},\set{C} \in \lattice{T}$ such that $\set{A} \sqsubseteq \set{B} \sqsubseteq \set{C}$ and $f \in \opt{O}{A}$ and $g \in \opt{O}{C}$ such that $\lhom{A}{C}(f) \sqsubseteq g$, then $\lhom{A}{B}(f) \sqsubseteq \lhom{C}{B}(g)$.
\end{theorem}
\begin{IEEEproof}
	Take arbitrary $t \in \set{B}$.

	\noindent \emph{Case 1}: $t \in \set{C}$.
	Then $f(t) \subseteq g(t)$, $\lhom{A}{B}(f)(t) = f(t)$, and $\lhom{C}{B}(g)(t) = g(t)$.
	Therefore $\lhom{A}{B}(f)(t) \subseteq \lhom{C}{B}(f)(t)$.

	\noindent \emph{Case 2}: $t \not\in \set{C}$.
	Then $\lhom{C}{B}(g)(t) = \set{O}$ and $\lhom{A}{B}(f)(t) \subseteq \set{O}$.
\end{IEEEproof}

\subsection{Tag-Options Lattice}

So far, we have developed a lattice that relates sets of tags that apply to a given system
and a family of lattices that relate option values given a set of tags.
Now we combine these into a lattice that can relate option values between sets of tags.

\begin{definition}
	A \emph{tag--options lattice} $\tol{T}{O}$ is a lattice of tuples $(\set{T'},f)$ where $\set{T'}$ is an element of $\lattice{T}$ and $f$ is an element of $\opt{O}{T'}$.
	Given elements $(\set{A},f)$ and $(\set{B},g) \in \tol{T}{O}$, $(\set{A},f) \sqsubseteq (\set{B},g)$ if and only if $\set{A} \sqsubseteq \set{B}$ and $\lhom{A}{B}(f) \sqsubseteq g$.
	
	(Note that if $(\set{A}, f) \sqsubseteq (\set{B},g)$, $\set{A} \sqcup \set{B} = \set{B}$, so $\lhom{A}{A \sqcup B} = \lhom{A}{B}$ and $\lhom{B}{A \sqcup B} = \lambda x.x$.)

	For any subset $\set{\Lambda} \subseteq \tol{T}{O}$, let $\set{V} \defeq \{T' \mid (T',f) \in \set{\Lambda}\}$.
	Then we can define
	\begin{enumerate}[i)]
		\item $\bigsqcup \set{\Lambda} \defeq \left(\bigsqcup \set{V},\bigsqcup\left\{\lhom{T'}{\bigsqcup \set{V}}(f) \mid (T',f) \in \set{\Lambda}\right\}\right)$, and
		\item $\bigsqcap \set{\Lambda} \defeq \left(\bigsqcap \set{V},\bigsqcap\left\{\lhom{T'}{\bigsqcap \set{V}}(f) \mid (T',f) \in \set{\Lambda}\right\}\right)$.
	\end{enumerate}
\end{definition}

In this lattice, $\bot = (\bot_{\lattice{T}}, \bot_{\opt{O}{T}})$ is the system where no options are valid for any tag.
$\top = (\top_{\lattice{T}}, \top_{\opt{O}{\emptyset}})$ corresponds to the system where no tags are known to apply.

For example, suppose we have a set of tags $\set{T} = \{t_1,t_2,t_3\}$ and take elements $\set{T_1}, \set{T_2} \in \lattice{T}$ where $\set{T_1} = \{t_1,t_2\}$, and $\set{T_2} = \{t_2,t_3\}$.
Let the options set be $\set{O} = \{x,y\}$.

Let $f : \set{T_1} \rightarrow \powerset{\set{O}}$ be an element of $\opt{O}{T_1}$ where $f = \{(t_1, \emptyset), (t_2, \set{O})\}$.
Let $g : \set{T_2} \rightarrow \powerset{\set{O}}$ be an element of $\opt{O}{T_2}$ where $g = \{(t_2, \{x\}), (t_3, \{y\})\}$.
Then $(T_1,f)$ and $(T_2, g)$ are elements of $\tol{T}{O}$.

Furthermore, we can compute the meet of $(T_1, f)$ and $(T_2, g)$ as follows:
\begin{align*}
	& (\set{T_1},f) \sqcap (\set{T_2},g)\\
	=& (\set{T_1} \sqcap \set{T_2}, \lhom{T_1}{T_1 \sqcap T_2}(f) \sqcap \lhom{T_2}{T_1 \sqcap T_2}(g)) \\
	=& (\set{T_1} \cup \set{T_2}, \{(t_1, \emptyset), (t_2, \set{O}), (t_3, \set{O})\} \\
	&\qquad \sqcap \{(t_1, \set{O}), (t_2, \{x\}), (t_3, \{y\})\})\\
	=& (\set{T}, \{(t_1, \emptyset \cap \set{O}), (t_2, \set{O} \cap \{x\}), (t_3, \set{O} \cap \{y\})\}) \\
	=& (\set{T}, \{(t_1, \emptyset), (t_2, \{x\}), (t_3, \{y\})\}).
\end{align*}

\begin{theorem}
	$\tol{T}{O}$ is a complete lattice.
\end{theorem}
\begin{IEEEproof}
	That $\sqsubseteq$ is a partial order follows from the partial orders defined on $\lattice{T}$ and the lattices of $\ofam{O}{T}$.

	To show that $\bigsqcap$ is a well--defined meet operator, take an arbitrary subset $\set{\Lambda} \subseteq \tol{T}{O}$, and let $\set{V} \defeq \{T \mid (T,f) \in \set{\Lambda}\}$.

	Take arbitrary $(T', f') \in \tol{T}{O}$ such that $\forall l \in \set{\Lambda}, (T',f') \sqsubseteq l$.
	Thus, $T' \sqsubseteq \bigsqcap \set{V}$.
	Furthermore,
	\begin{align*}
		  & \forall l \in \set{\Lambda}, (T', f') \sqsubseteq l \\
		\implies
		  & \forall (T, f) \in \set{\Lambda}, T' \sqsubseteq T \wedge \lhom{T'}{T}(f') \sqsubseteq f
		  & (\text{Defn. of $\sqsubseteq$})
		  \\
		\implies
		  & \forall (T, f) \in \set{\Lambda}, \lhom{T'}{\bigsqcap \set{V}}(f') \sqsubseteq \lhom{T}{\bigsqcap \set{V}}(f)
		  & (\text{Thm.~\ref{thm:lhom-order-corr}})
		  \\
		\implies
		  & \lhom{T'}{\bigsqcap \set{V}}(f') \sqsubseteq \bigsqcap \left\{\lhom{T}{\bigsqcap\set{V}}(f) \mid (T, f) \in \set{\Lambda}\right\} \\
		  \noalign{\raggedleft $($Defn. of $\bigsqcap$ for $\opt{O}{\bigsqcap \set{V}}$$)$}
	\end{align*}
	Therefore $(T', f') \sqsubseteq \bigsqcap \set{\Lambda}$.
	The proof that $\bigsqcup$ is a well--defined join operator is analogous.
\end{IEEEproof}


For an example of how a tag--options lattice might be used to construct a concrete properties domain, consider the task of assigning reliabilities to components.
In this case, we define a set of component names $\set{Components} \defeq \{c_1, c_2, \cdots\}$ to use as tags.
The elements of the tag lattice $\lattice{Components} \defeq \powerset{\set{Components}}^{\partial}$ consist of sets of component names.
If an element of $\lattice{Components}$ applies to a given system, then we know that the system consists of at least those components.
The set of options is $\set{Probability} = \{x \in \mathbb{R} \mid 0 < x \leq 1\}$.
Finally, the tag--options lattice $\tol{Components}{Probability}$ consists of pairs $(\set{C}, p)$ where $\set{C}$ is a set of components known to comprise a given system and $p(c)$ assigns a range of possible reliabilities to each $c \in \set{C}$.
Thus $\tol{Components}{Probability}$ is a lattice of constraints on component reliability ordered by specificity.
It can be used as part of a definition of a properties domain in conjunction with other lattices that capture other relevant system properties.

\section{Conclusion and Future Work}
\label{sec:conclusion}

In this paper, we have demonstrated a formalization of model and system semantics.
Models abstract system semantics; therefore, we can derive, or concretize, constraints on a system from models of it.
Conversely, given constraints on a system, we can abstract a set of models that are consistent with those constraints.

To formalize the soundness of this approach, we apply abstract interpretation, which defines a correctness relation
between systems and constraints.
If our abstraction and concretization mappings between a given modeling formalism and system constraints form a Galois connection between the two domains, we can show that these mappings and the correctness relation for system constraints induce a correctness relation between systems and the models of the modeling formalism.

Through this lens, the process of model transformation becomes the process of concretizing system properties from one model, then abstracting a second model from these properties.
We show that this process is sound; that is, if the initial model is correct, then the final model will also be correct.

Future work will take several directions.
We are currently working on relating models of different aspects of a system---in this case, reliability and topology.
This work will demonstrate both how topology affects system reliability by introducing dependencies between components
and how reliability, via the same dependencies and the constraints on system functionality,
constrains the choice of topologies for which that definition of reliability holds.

We plan to further extend the work of this paper to other model types and other choices of system properties.
Expanding the possible transformations will allow us to relate modeling techniques from various system domains;
for example, we may apply a water distribution network analysis technique to a power grid,
or incorporate both cyber and physical models into a cyber--physical model.

Another avenue of research is to expand the formalization of models and systems to other metamodeling tasks.
A salient challenge in the design of complex systems is that of heterogeneous model composition: combining component models
that use various modeling formalisms into a single model of a complete system.
The abstraction and concretization functions defined in this work provide a basis for developing these connections.
It may even be possible to perform this composition at a higher level, enabling the creation of hybrid modeling formalisms and associated solution and evaluation techniques.

Finally, the task of developing this approach into a tool for system designers will certainly present its own challenge.
Such a tool must be interactive and scalable to complex, real--world systems, all without requiring the user to have a deep understanding of the underlying theory.


\bibliographystyle{ieeetr}
\bibliography{references,lit-review}

\end{document}